\documentclass[twocolumn,prl]{revtex4}
\usepackage{graphics,graphicx}
\begin{document}
\title{Temperature Dependent Magnetism in Artificial Honeycomb Lattice of Connected Elements}
\author{B. ~Summers$^{1}$}
\author{L. Debeer-Schmitt$^{2}$}
\author{A. Dahal$^{1}$}
\author{A. Glavic$^{2,3}$}
\author{P. Kampschroeder$^{1}$}
\author{J. Gunasekera$^{1}$}
\author{D. ~K.~Singh$^{1,*}$}
\affiliation{$^{1}$Department of Physics and Astronomy, University of Missouri, Columbia, MO 65211}
\affiliation{$^{2}$Oak Ridge National Laboratory, Oak Ridge, TN 37831}
\affiliation{$^{3}$Laboratory for Neutron Scattering and Imaging, Paul Scherrer Institut, Villigen PSI, Switzerland}
\affiliation{$^{*}$email: singhdk@missouri.edu}

\begin{abstract}
Artificial magnetic honeycomb lattices are expected to exhibit a broad and tunable range of novel magnetic phenomena that would be difficult to achieve in natural materials, such as long-range spin ice, entropy-driven magnetic charge-ordered state and spin-order due to the spin chirality. Eventually, the spin correlation is expected to develop into a unique spin solid state density ground state, manifested by the distribution of the pairs of vortex states of opposite chirality. Here we report the creation of a new artificial permalloy honeycomb lattice of ultra-small connecting bonds, with a typical size of $\simeq$ 12 nm. Detail magnetic and neutron scattering measurements on the newly fabricated honeycomb lattice demonstrate the evolution of magnetic correlation as a function of temperature. At low enough temperature, neutron scattering measurements and micromagnetic simulation suggest the development of loop state of vortex configuration in this system.\end{abstract}

\pacs{75.75.-c, 75.90.+w, 75.25.-j, 85.75.-d}

\maketitle

\textbf{1. Introduction}  

Artificial magnetic honeycomb lattice manifests a two dimensional prototype of three dimensional geometrically frustrated magnets where intriguing, yet, novel magnetism has been intensively explored in recent times.\cite{Nisoli,Ramirez} It includes the ice analogue of magnetism, spin ice, spin liquid and exotic quantum mechanical properties of the resonant valence bond state.\cite{Gardener,Snyder,Harry} The concept of an artificial honeycomb lattice or a two dimensional artificial structure was originally conceived to study the physics of spin ice state.\cite{Tanaka,Wang,Mengotti,Qi} Since then it has evolved into a general arena to not only explore the entire spectrum of the novel magnetism in geometrically frustrated magnet but also a broad and tunable range of magnetic phenomena that would be difficult to achieve in natural materials.\cite{Branford,Olson} It became possible due to a recent theoretical proposal, which suggests that a magnetic moment or spin can be considered as a pair of magnetic charges of opposite polarities, as if it is a 'dumbbell', that interact via the Coulomb interaction.\cite{Sondhi1,Fennell} The direction of magnetic moment or spin points from the negative to the positive charge.

The concept of magnetic charge can be utilized to describe the competing magnetic states in artificial honeycomb lattice. Under this scheme, the moment along the honeycomb element can be represented as a dipole of $\pm$1 unit charge. Since each vertex of the honeycomb lattice is joined by three moments, it can possess a net magnetic charge of $\pm$3 or $\pm$1 unit depending on the direction of the moment along the honeycomb element (as described schematically in Fig. 1a-c).\cite{Ladak} Here, charges $\pm$3 arise when all three magnetic moments along the honeycomb element point to or away from the vertex at the same time, respectively. Such arrangement of moments is called 'all-in or all-out' configuration. On the other hand, if two of the moments point to the vertex and one points away from the vertex (or vice-versa), then the vertex possesses a net magnetic charge of $\pm$1 unit. This is often referred to as 'two-in \& one-out' (or vice-versa) configuration. At sufficiently high temperature, the lattice can be described as a paramagnetic gas, consists of the random distribution of $\pm$3 and $\pm$1 magnetic charges. Recent theoretical calculations have shown that an artificial magnetic honeycomb lattice can undergo a variety of novel ordered regimes of correlated spins and magnetic charges of both fundamental and practical importance as a function of temperature. It includes long-range spin ice, entropy-driven magnetic charge-ordered state and spin-order due to the spin chirality as a function of reducing temperature.\cite{Moller,Chern,Mellado,Crespi} At low enough temperature, magnetic correlation is expected to develop into a spin solid state density where the magnetization profile assumes a chiral vortex configuration involving six vertexes of the honeycomb lattice (see Fig. 1d).\cite{Rougemaille1} The spin solid state, manifested by the distribution of the pairs of vortex states of opposite chiralities across the lattice, provides a unique opportunity to realize a magnetic material with net zero entropy and magnetization for an ordered ensemble of magnetic moments.\cite{Branford, Rougemaille1}

\begin{figure*}
\centering
\includegraphics[width=18 cm]{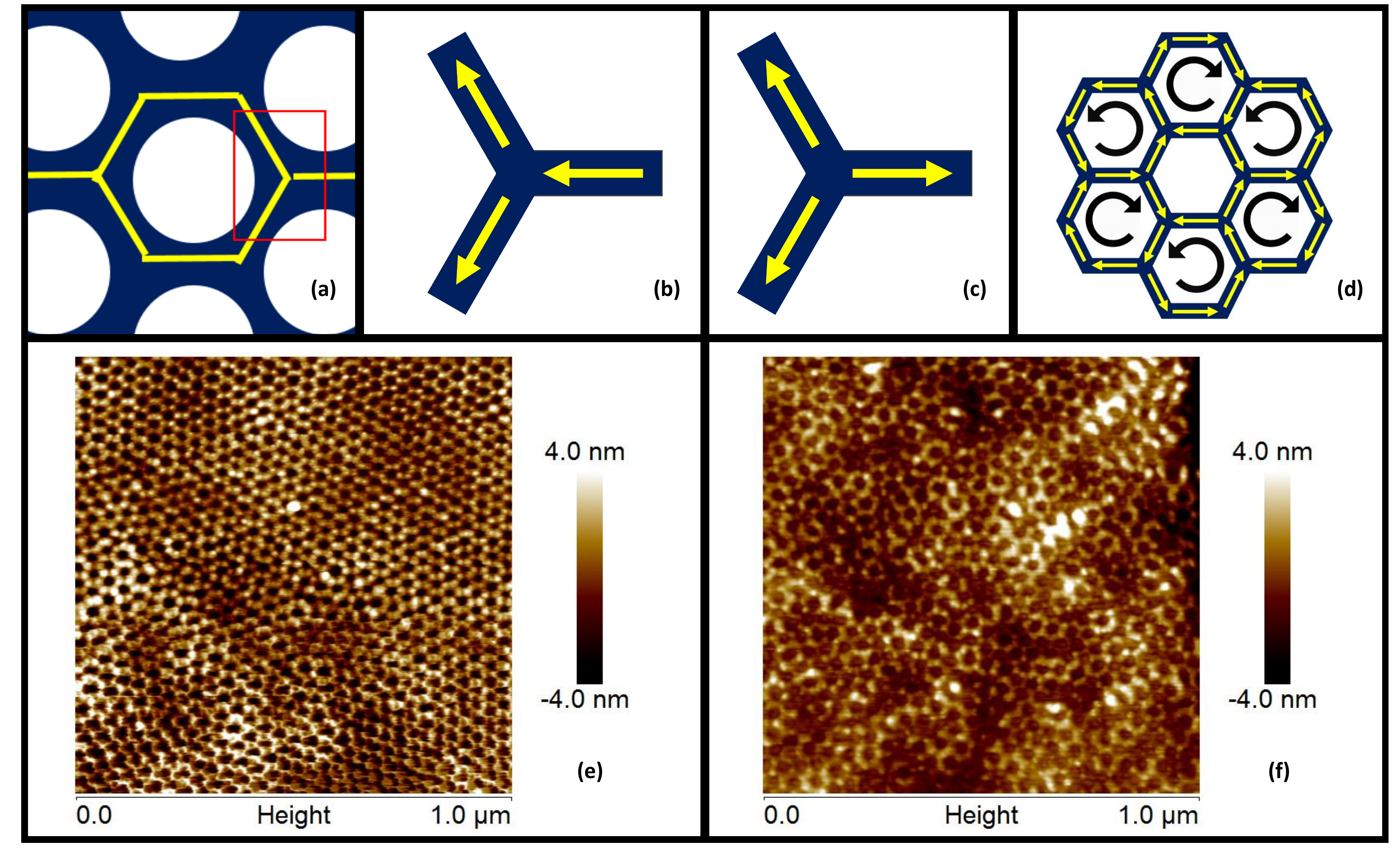} \vspace{-4mm}
\caption{Schematic description of spin configuration on a two dimensional honeycomb lattice vertex and the atomic force micrograph. (a-c) Typical 2-out 1-in and all-out spin arrangements on a vertex of a 2-dimensional artificial magnetic honeycomb lattice resulting into net magnetic charges of $\pm$Q and $\pm$3Q, respectively. (d) Theoretical researches suggest that at sufficiently low temperature, magnetic charges at the vertexes of an artificial honeycomb lattice can arrange themselves to create a spin solid state, manifested by the periodic arrangement of pairs of chiral vortex states. (e) A full size atomic force micrograph of artificial honeycomb lattice, derived from diblock porous template combined with reactive ion etching. The bond length, width and lattice separation are approximately 12 nm, 5 nm and 26 nm, respectively. (f) Atomic force micrograph of a typical metallic honeycomb lattice fabricated using the method described in the text and the Supplementary Materials.} \vspace{-4mm}
\end{figure*}

The experimental efforts to realize the temperature dependent magnetic correlations in an artificial honeycomb lattice is limited due to the employment of the present nanofabrication method of electron-beam lithography (EBL).\cite{Nisoli} The EBL technique results in small sample size with large connecting element (or bond of the honeycomb lattice), of the order of 500 nm - few $\mu$m. Such a large element size leads to the inter-elemental energy of 10$^{4}$-10$^{5}$ K.\cite{Nisoli,Rougemaille2}. Recent explorations of large element size disconnected honeycomb lattice and other frustrated two-dimensional structures e.g. square spin ice or tetris lattice have revealed interesting thermal dependence of the magnetic properties.\cite{Zhang,Poro,Gilbert,Farhan} An alternative approach to reduce the inter-elemental energy involves the idea of a disconnected honeycomb structure where magnetic elements are very thin and well-separated. It has been shown that such a sample fabrication design significantly reduces the inter-elemental energy between the honeycomb bonds, thus makes it possible to explore the predicted temperature dependent phase diagram.\cite{Heyderman1,Heyderman2,Heyderman3,Arnalds}

Here we propose a new nanofabrication scheme, which allows the creation of macroscopic size artificial honeycomb lattice with ultra-small dimension of the connecting elements, $\simeq$ 12 nm$\times$5 nm$\times$5 nm. We report the temperature dependence of magnetization in the connected artificial honeycomb lattice, which exhibits a near zero magnetization at low temperature. The ultra-small connecting elements of the honeycomb lattice exhibit an estimated dipolar energy of the order of $\simeq$ 10 K. Such a small inter-elemental energy makes the application of temperature as a feasible tuning parameter to explore novel magnetic correlations in the artificial magnetic honeycomb lattice. Magnetic and neutron scattering measurements on the newly fabricated system indeed reveal multiple magnetic regimes, suggestive of varying magnetic correlations, as a function of the reducing temperature. At low temperature, $T \leq$ 30 K, neutron scattering measurements suggest that the system tends to develop loop state of the vortex configuration. 

\textbf{2. Sample Fabrication and Measurement Methods} 

The fabrication scheme of new artificial honeycomb lattice is described in detail in the Supplementary Materials. Basically, it begins with the creation of a nanoporous polymer template, derived from a self-assembled diblock copolymer film, on top of a silicon substrate (see Fig. S1 in the Supplementary Materials).\cite{Supplemental,Park} The nanoporous template exhibits a typical pore diameter of $\simeq$ 12 nm and the center-to-center lattice spacing of 26 nm. The silicon substrate, underneath the diblock template, is reactively etched using CF$_{4}$ gas to transfer the hexagonal pattern to the substrate. In Fig. 1e, we show the atomic force micrograph of the resulting hexagonal structure in silicon substrate. The top layer of the hexagonal substrate depicts a honeycomb pattern. This property is exploited to create a two-dimensional metallic honeycomb lattice of the ultra-small bond by depositing magnetic material in almost parallel orientation on top of the hexagonal silicon substrate only (see Fig. S2 in the Supplementary Materials for a schematic description).\cite{Supplemental} The substrate is uniformly rotated about its axis during the deposition process. Fig. 1f shows the AFM image of a typical metallic honeycomb lattice. The new fabrication method, utilizing the diblock template technique, provides a large throughput sample of the artificial honeycomb lattice, which is quite suitable for the bulk properties investigation using various macroscopic probes. The honeycomb samples were preserved in vacuum environment to reduce the exposure to air. Magnetic measurements were performed using a QD MPMS on a 5 mm $\times$ 5 mm size sample. Polarized neutron scattering experiments were performed on a 1 sq. inch sample at magnetism reflectometer, beamline 4A of the Spallation Neutron Source (SNS), at Oak Ridge National Laboratory. The instrument used the time of fight technique in a horizontal scattering geometry. The beam was collimated with a set of slits before the sample and measured with a 2D position sensitive $^{3}$He detector. Polarization and analysis used reflective supermirror technology.

\textbf{3. Magnetic and Neutron Scattering Measurements}

Magnetometry is a key macroscopic probe to obtain information about the static and dynamic magnetic properties of a system as functions of temperature and field. The macroscopic size of the newly designed artificial honeycomb lattice is well suited for investigation using this measurement technique. We have performed detailed magnetic measurements on the recently fabricated artificial honeycomb lattice of connecting permalloy (Ni$_{0.8}$Fe$_{0.2}$) bonds. Magnetic field was applied along an in-plane direction to the sample.  As shown in Fig. 2a, the ZFC (zero field cool) / FC (field cool) curves of $M$ vs. $T$ measurements depict the temperature dependence with multiple magnetic regimes in the honeycomb lattice. At T $\geq$ 300 K, the system is a paramagnetic gas (spin gas). As temperature is reduced, the system crosses over into a weak magnetic ordered state at $T$$\simeq$ 250 K, indicated by small downward cusp in the low field data (also see the inset in Fig. 2a). For further decrease in temperature below $T$ = 100 K, another small downward cusp-- indicating a new magnetic regime-- is detected. As the applied field increases, the irreversibility between the FC and ZFC curves gradually shifts to lower temperature, before disappearing at $H$ $\geq$ 500 Oe. The strong sensitivity of various magnetic correlation regimes to the applied field is also consistent with previous observation of the field-induced avalanche effect in the large element size artificial honeycomb lattice, where the field application tends to destroy the delicate short-range spin ice order due to 2-in \& 1-out (or vice-versa) magnetic configuration. \cite{Nisoli,Mengotti,Ladak} 

\begin{figure*}
\centering
\includegraphics[width=15 cm]{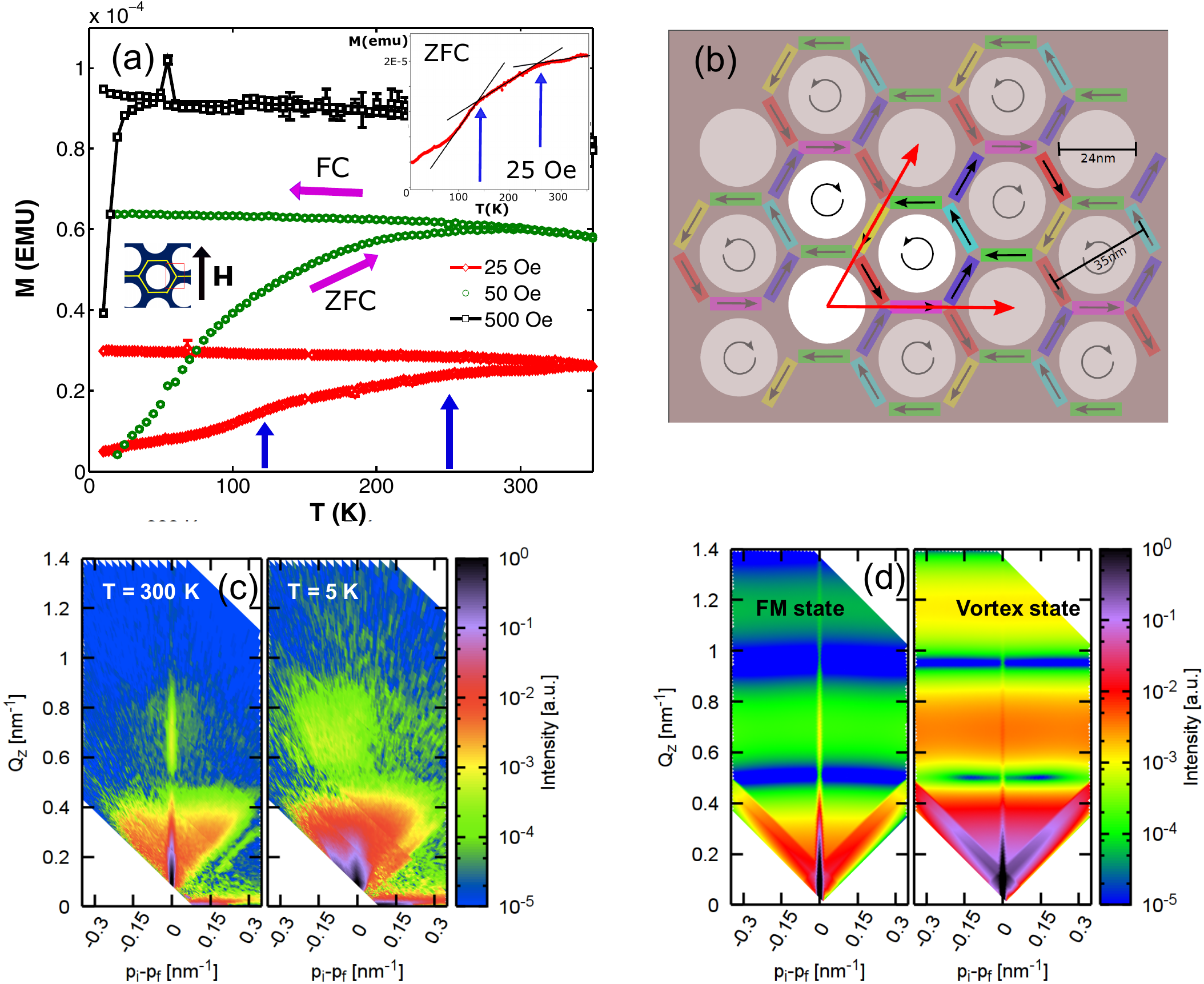} \vspace{-4mm}
\caption{In-plane magnetic measurements and polarized off-specular neutron reflectometry data. (a) Here we show M vs. T data in zero field cool (ZFC) and field cool (FC) measurements at characteristic fields. M vs T measurements in different fields not only exhibit strong temperature dependence but also reveal multiple ordering regimes as functions of field and temperature (also see inset). (b) Spin solid state configuration, used to simulate the off-specular polarized reflectometry profile (Fig. d (right)). (c) Off-specular neutron reflectometry data recorded with spin-up incident polarization at T = 300 K and 5 K, respectively. Here, the x-axis indicates the inplane correlation while the y-axis indicates the out-of-plane correlation. The specular reflection at room temperature, indicating paramagnetic state, is replaced by a broad diffuse scattering extending along the x-axis, primarily due to the development of the spin solid phase. (d) Numerically simulated off-specular neutron reflectometry profiles for paramagnetic (weakly FM) honeycomb film (left) and the spin solid state (right) are consistent with the experimental data. Compared to the off-specular data where very weak intensity is observed, specular reflection is strong in the FM case (left). Unlike the FM case, simulated profile using the spin solid state (as shown in Fig. b) exhibits bands of broad scattering along the horizontal axis (right) with almost negligible specular intensity. Error bar represents one standard deviation in the experimental data.
} \vspace{-4mm}
\end{figure*}

The temperature dependent magnetization curves exhibit a tendency to attend the zero magnetization state (see ZFC curves) at temperature below $T$ = 30 K. This behavior becomes more apparent at higher magnetic field. For instance, at $H$ = 500 Oe, the net magnetization of the honeycomb lattice reduces rapidly towards zero value at $T$$\leq$30 K from the large saturation value. This behavior is only observed in the zero field cool measurement i.e. when the sample is cooled to the base temperature in zero magnetic field. Thus, the system develops the near zero magnetization state in the 'absence' of magnetic field. As soon as a magnetic field is applied, the correlated moments tend to abandon that delicate zero magnetization state. When cooled back in applied field (as small as $H$ = 25 Oe), the moments remain locked in to the field-aligned value. The lock-in temperature reduces with increasing magnetic field. Our efforts of accessing the net zero magnetization state in the newly fabricated permalloy honeycomb lattice was hampered by the technical limitation of the present SQUID magnetometer, which could not be cooled below $T$ = 5 K. Nonetheless, the trend towards zero moment as $T$$\rightarrow$ 0 K is apparent in the ZFC curve of the magnetization data. Magnetic measurements were also performed for the perpendicular field application to the sample plane. As shown in Fig. S3 in the Supporting Materials,\cite{Supplemental} no appreciable change in the magnetization pattern of ZFC and FC curves were detected for the perpendicular field application. This behavior is not surprising for such a thin ($\simeq$ 5 nm) honeycomb film.\cite{Heyderman} The perpendicular direction acts as the hard axis for the magnetization reversal to take place.

\begin{figure*}
\centering
\includegraphics[width=17.8 cm]{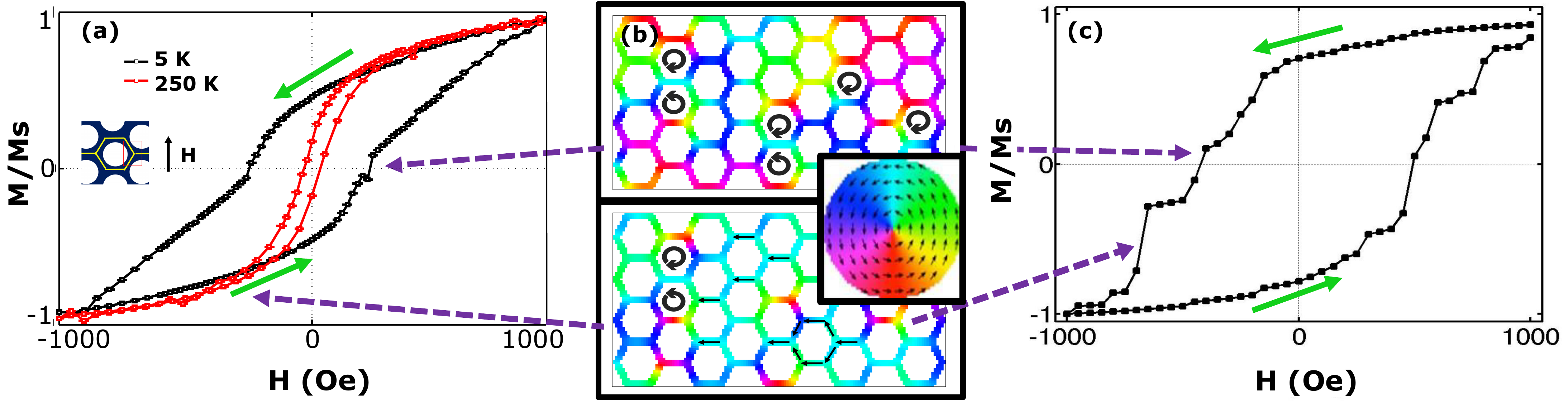} \vspace{-4mm}
\caption{Magnetic measurements in applied magnetic field and micromagnetic simulation results of permalloy honeycomb lattice. (a) M vs. H measurements at two different temperatures. Magnetic field was applied in-plane to the sample. Error bar represents one standard deviation in the experimental data. (b) Experimental data is compared with micromagnetic simulation. Near zero field, the artificial honeycomb lattice exhibits 2-in \& 1-out spin ice structure. As the field is reduced, pair of chiral vortexes arise as highlighted by the arrow. (c) The magnetic hysteresis loop obtained from micromagnetic simulation agrees well with the experimental data. The micromagnetic simulation was performed using 0.2 $\times$ 0.2 nm$^{2}$ mesh size, with magnetic field applied in-plane to the lattice. 
} \vspace{-4mm}
\end{figure*}

In order to gain more insight in the low temperature magnetic properties in artificial permalloy honeycomb lattice of ultra-small elements, we have also performed polarized neutron experiments, namely reflectometry (PNR) and off-specular scattering. The off-specular measurements allow us to understand the development of the in-plane magnetic structure as a function of temperature in the system. In Fig. 2c, we plot the off-specular data in the spin up polarization channel at $T$ = 300K and 5K where the vertical direction corresponds to the out-of-plane correlation and the horizontal to the in-plane correlation. The vertical line across the origin represents the specular reflectivity. The measurement at $T$ = 300 K already exhibits significant intensity in the specular data, as is typical for most samples and can be expected due to the saturated honeycomb structure with no inplane magnetic contrast. Upon cooling to $T$ = 5 K, the off-specular signal increases significantly (notice the logarithmic color scale). Also, no specular beam can be distinguished from the off-specular background and the difference between the spin-up and the spin-down component vanishes. As the nuclear structure will not change significantly upon cooling, this can only be explained by a significant change in the magnetic order. The signal itself is very flat along the x-direction, suggesting the development of an in-plane magnetic correlation. Numerical simulation of the scattering profile, see Fig. 2d, using a vortex magnetic configuration of the spin solid state, as shown in Fig. 2b, reproduces essential features of the experimental data, such as the band of broad scattering along the horizontal direction and an almost negligible specular reflection. It suggests that the system tends to develop a spin solid state at low temperature.

The underlying magnetism at low temperature is further investigated using magnetic hysteresis measurements. In Fig. 3a, we plot $M$ vs $H$ data at two characteristic temperatures of $T$ = 5 K and 250 K. Measurement at $T$ = 5 K reveals a sharp transition to a near zero magnetization state near the zero field value, which is completely absent at $T$ = 250 K. To understand this, we have performed micromagnetic simulations on artificial permalloy honeycomb lattice of similar element size and thickness by utilizing the Landau-Lifshitz-Gilbert equation of magnetization relaxation in a damped medium.\cite{Brown} The artificial honeycomb lattice was simulated using 0.2 $\times$ 0.2 nm$^{2}$ mess size on the OOMMF platform, with magnetic field applied in-plane to the lattice.\cite{OOMMF} The simulated magnetic hysteresis curve is shown in Fig. 3b, which depicts striking similarity with the experimental data. The magnetic correlation near zero field is found to be dominated by the distribution of the chiral vortex configurations. At moderate field value, the finite magnetization in the artificial honeycomb lattice is described by the short-range spin ice correlation of 2-in \& 1-out (and vice-versa) states. At sufficiently high field, the moments tend to align to the applied field direction, thus maximizing the overall magnetization of the system. The micromagnetic simulations were also performed for the honeycomb lattice with distorted bond dimensions, varying between 10-15 nm in length, 4-7 nm in width and 4-7 nm in thickness, to understand the role of the quenched disorder in the system. As shown in Fig. S5 in the Supplementary Materials, no significant change in the magnetic hysteresis profile was detected in this case.\cite{Supplemental}

\textbf{4. Discussion} 

In summary, we have presented a novel fabrication scheme to create macroscopic size artificial honeycomb lattice of ultra-small element. Detailed measurements on the newly fabricated permalloy honeycomb lattice reveal the temperature dependent evolution of magnetic correlation in this two dimensional geometry. While the analysis of neutron data suggests the development of the loop state of spin solid phase at low temperature, there are several questions that remain unanswered. For instance, the mechanism behind the zero magnetization state in ZFC/FC measurement is not understood. Although, the system is expected to develop a zero magnetization state in the spin solid state as $T \rightarrow$ 0 K, but it can also arise due to other mechanism, such as the random distribution of moments. Therefore, further research works are needed to understand the origin of zero magnetization state. The proposed sample design can also be helpful in exploring temperature dependent development of Dirac string due to the effective monopoles.\cite{Nisoli} Previously, such a state was demonstrated to exist in applied magnetic field in large element size magnetic honeycomb lattice using X-ray dichroism measurements. \cite{Mengotti}

\textbf{Acknowledgements} 

We acknowledge helpful discussion with Giovanni Vignale. The research at MU was supported by the U.S. Department of Energy, Office of Basic Energy Sciences under Grant No. DE-SC0014461. A portion of this research used resources at the Spallation Neutron Source, a DOE Office of Science User Facility operated by the Oak Ridge National Laboratory. BS acknowledges the NSF IGERT fellowship at MU under grant number DGE-1069091.

\clearpage

\end{document}